\documentclass[
aps,%
12pt,%
final,%
notitlepage,%
oneside,%
onecolumn,%
nobibnotes,%
nofootinbib,
superscriptaddress,%
noshowpacs,%
centertags]%
{revtex4}

\begin{document}
\selectlanguage{english} 
\title{High p$_{T}$ and jet physics from RHIC to LHC}
\author{\firstname{M.} \surname{Estienne}} 
\email[]{magali.estienne@subatech.in2p3.fr}
\affiliation{Institut Pluridisciplinaire Hubert Curien, Strasbourg, FRANCE}
\collaboration{for the ALICE Collaboration}
\begin{abstract}
The observation of the strong suppression of high $p_{T}$ hadrons in
heavy ion collisions at the Relativistic Heavy Ion Collider (RHIC) at
BNL has motivated a large experimental program using hard probes to
characterize the deconfined medium created. However
what can be denoted as ``leading particle physics'' accessible at RHIC
presents some limitations which motivate at higher energy the study of
much more penetrating objects: jets. The gain in center of
mass energy expected at the Large Hadron Collider (LHC) at CERN will
definitively improve our understanding on how the energy is lost in the
system opening a new major window of study: the physics of jets on an
event-by-event basis. We will concentrate on the expected performance
for jet reconstruction in ALICE using the EMCal
calorimeter. \\
\end{abstract}
\maketitle
\section{Introduction}
\label{sec:I}
Following an initial hard scattering in $e^{+}+e^{-}$, $e^{-}+p$ and
hadron collisions, high energetic partons create one (or more) high
energy cluster(s) of particles moving in a same collimated
direction. These kind of global objects are called ``jets''. Parton
showering and the subsequent hadronization which lead to the particle
production are known as ``parton fragmentation''. In
ultra-relativistic heavy ion collisions (HIC), the scene of parton
fragmentation is changed from vacuum to a QCD medium. These partons
created before the QGP formation will first travel through the dense
color medium before the ``jet'' formation. They are expected to lose
energy through collisional energy loss and medium induced gluon
radiation~\cite{9}. This is typically known as {\it jet quenching}. The magnitude
of the energy loss depends on the gluon density of the medium
($dN_{g}/dy$) or on the number of scattering centers usually
quantified with $\hat{q}$ the transport coefficient and on its pass
length squared~\cite{10}. 

The measurement of the parton fragmentation products thus may yield
information about the QCD medium. Even if the total jet energy is
conserved, the phenomenon of quenching should change the jet
fragmentation function and its structure. In the former, the
manifestation of the partonic energy loss is a decrease of the number
of particles carrying a high fraction of the jet energy
E$_{\hbox{\scriptsize jet}}$ (high $z = E_{\hbox{\scriptsize hadron}}
/ E_{\hbox{\scriptsize jet}}$ or low $\xi =
ln(\frac{1}{\hbox{\scriptsize z}}$)) and an increase of the number of
low-energy particles (low $z$ or high $\xi$) results of the radiated
energy. In the latter, a broadening of the distribution of
jet-particle momenta perpendicular to the jet axis, $j_{T}$, directly
related to the colour density of the traversed medium is
expected~\cite{11,12}. Such description would be the ideal scenario
for the study of jet quenching under the assumption that jet
measurement is possible in a heavy ion environment. Looking at RHIC
Au+Au collisions at $\sqrt{s_{NN}}$ = 200 GeV, one can estimate that
more than 350 GeV energy is included in a typical cone size 
$\hbox{R} = \sqrt{\Delta\eta^2+\Delta\phi^2} = 1$~\cite{13}, whereas a
similar calculation for Pb+Pb collisions at $\sqrt{s_{NN}}$ = 5.5 TeV
at LHC leads to an energy of the underlying event of 1.5--2.0
TeV~\cite{14}. With such numbers in mind, one easily understands that
``jets'' at RHIC are very complicated objects which make them
impossible to disentangle from the ``background''. At such energies,
hard scattering is better understood with single particle and few
particle correlation measurements~\cite{13} (section~\ref{sec:II}).

At LHC, the picture should be quite different as 1 - the multi-jet
production per event is not restricted to the mini-jet region $E_{T} <
$ 2 GeV but extended to 20 GeV, 2 - the expected jet rate at which
jets can be clearly distinguished from the underlying event is
high. Although some ALICE physics performances in Pb+Pb have been
presented during my talk, in this paper, we will concentrate in
section~\ref{sec:III} on the $p+p$ case to show some improvements that
will bring the EMCal calorimeter recently proposed as an ALICE
upgrade~\cite{7,8,12}.
 
\section{RHIC and the ``leading'' particle physics}
\label{sec:II}
Observables such as the nuclear modification factor $R_{AA}$ given by
the ratio of A+A to $p+p$ invariant yields scaled by the nuclear
geometry ($T_{AB}$) and two and three particle correlations have
unequivocally shown at RHIC evidence for a non negligeable interaction
of high energy partons with a dense color medium before
hadronization~\cite{15,2,3,4,5,6}. The use of the word {\it jet
quenching} for the phenomenon which appeared consequently is a bit
confusing as it is more a suppression of the production of particles
with high p$_{T}$ than a suppression of the ``jets'' themselves that
is observed. It is also usually the most energetical particle
(trigger) that is correlated to the associated remnants and not the
jet itself. Even if this ``leading particle'' physics has already
demonstrated its large potential to understand hard scattering
processes in HIC, it has some limitations that will be
briefly discussed below.

\subsection{$p+p$ Baseline and strong suppression at RHIC}

\begin{figure}[htbp]
\setcaptionmargin{5mm}
\onelinecaptionstrue
\includegraphics[scale=0.37, angle=0]{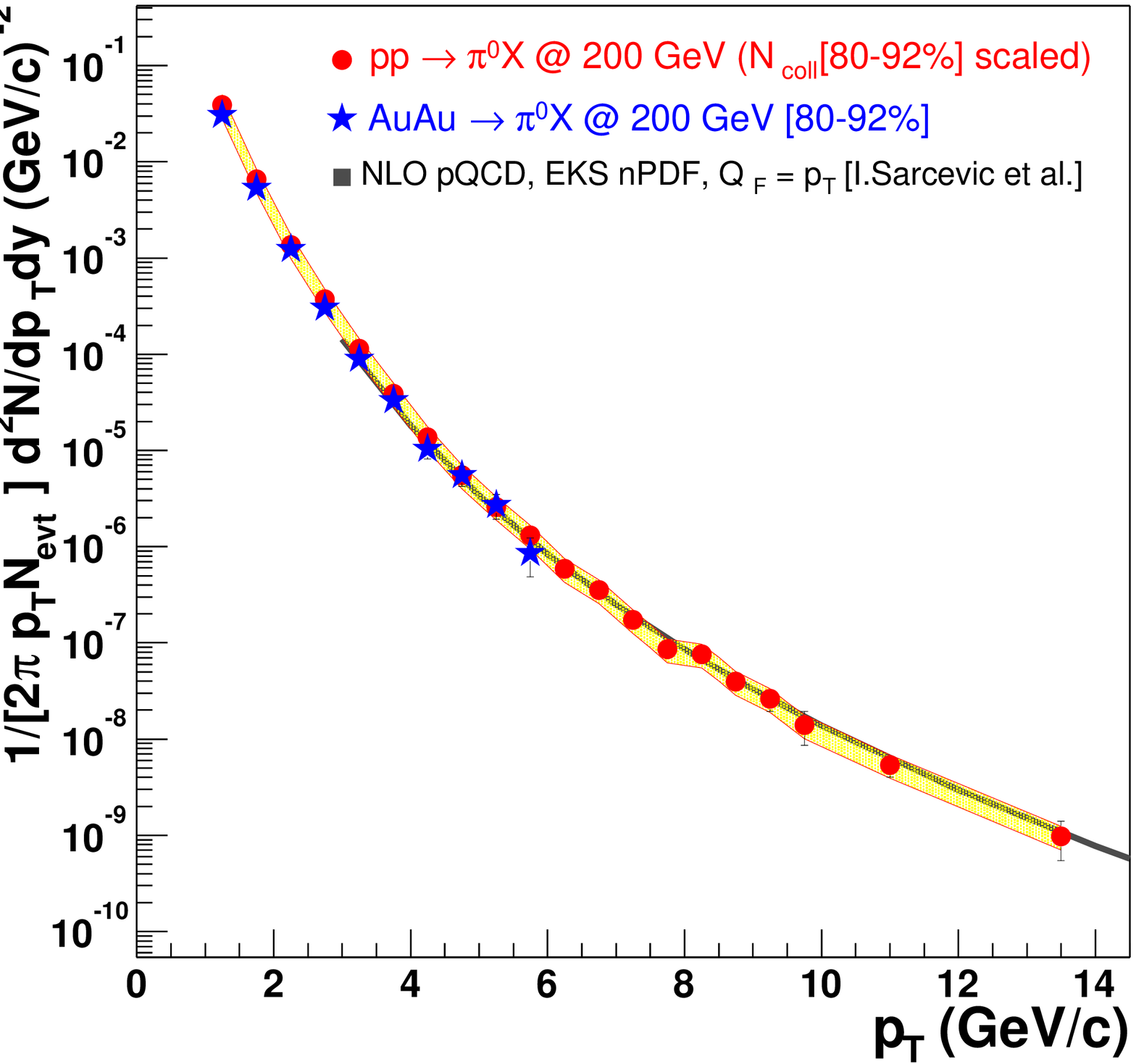}
\includegraphics[scale=0.37, angle=0]{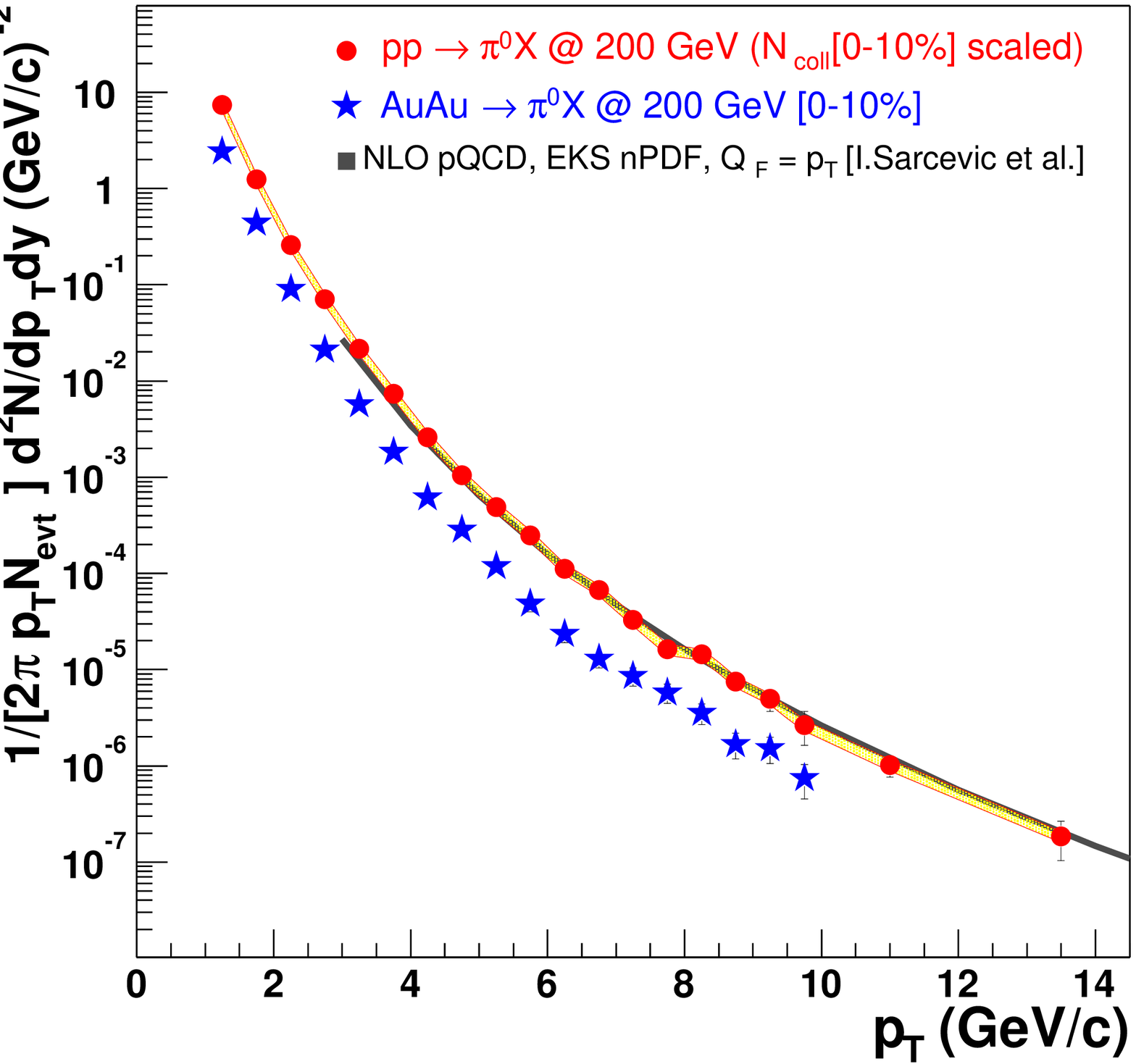}
\includegraphics[scale=0.37, angle=0]{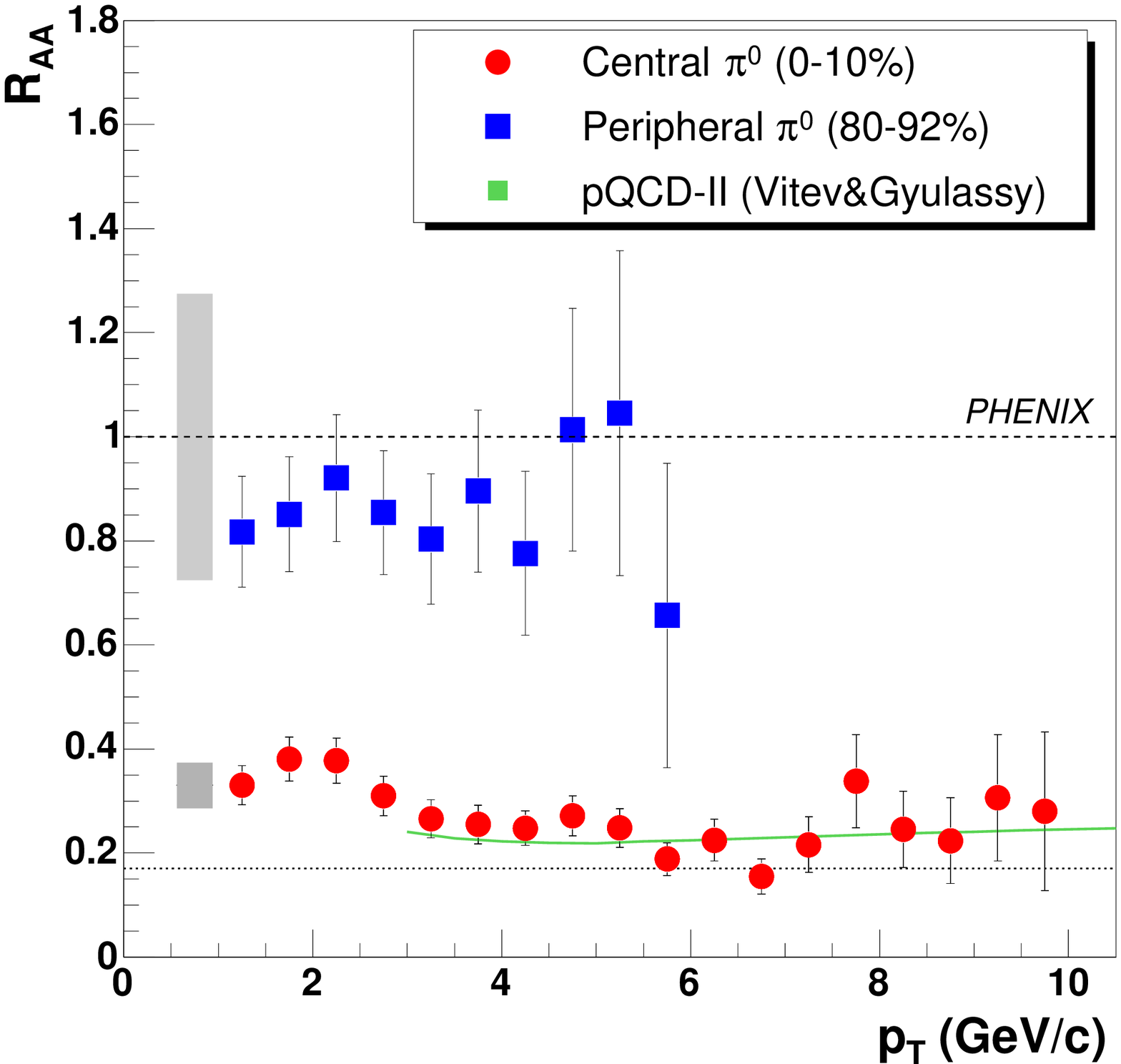}
\caption{{Top:} invariant $\pi^{0}$ yields measured by PHENIX in peripheral (left) and central (right) Au+Au collisions (stars), compared to the N$_{\hbox{\scriptsize coll}}$ $p+p$ scaled $\pi^{0}$ yields (circles) and to pQCD calculation (squares). {Bottom}: nuclear modification factor, $R_{AA}$, in peripheral and central Au+Au collisions at $\sqrt{s_{NN}}$ = 200 GeV for $\pi^{0}$ measured by PHENIX~\cite{15,16,17}.}
\label{fig:pi0}
\end{figure}

High-$p_{T}$ particle production in proton--proton collisions provides
the baseline ``vacuum'' reference to heavy-ion collisions to study the
QCD medium properties. It requires the physics of such collisions to
be well under control. This is now the case in the experiments STAR
and PHENIX. As an illustration, PHENIX results are displayed as they
presented a direct comparison of Au+Au to $p+p$
spectra. Figure~\ref{fig:pi0} (top) shows the invariant $\pi^{0}$
yields in $p+p$ collision scaled by $N_{\rm coll}$ for comparison to Au+Au
data~\cite{15,16}. One can see the really good agreement with a
standard Next--To--Leading (NLO) calculation~\cite{17}. Their published
measurement of high $p_{T}$ inclusive $\pi^{0}$ cross-section at
$\sqrt{s_{NN}}$ = 200 GeV is also well described by NLO perturbative
calculations~\cite{16}. A baseline well defined, the hadron production
mechanisms in N+N can be studied via their scaling behaviour with
respect to $p+p$ collisions. Figure~\ref{fig:pi0} shows the comparison
of the $p+p$ $\pi^{0}$ spectrum to peripheral (top-left) and central
(top-right) Au+Au collisions. Whereas the superposition of the $p+p$ scaled
and peripheral spectra suggests that peripheral collisions are nothing
but a superposition of nucleon--nucleon collisions, central
data exhibits a clear suppression of factor 4--5. As commented
previously, such observation is emphasized by looking at the ratio of
Au+Au over $p+p$ scaled spectra to build the nuclear modification
factor. The result of this exercise is shown in Fig.~\ref{fig:pi0}
(bottom) for peripheral $\pi^{0}$ (squares) and central ones
(circles). Within error bars, peripheral to $p+p$ scaled ratio is
consistent with unity, i.e. with a binary collision scaling. For
central data, the suppression smoothly increases with p$_{T}$ to a
constant suppression factor of 4--5. Within different model
interpretations, this observation would be consistent with a gluon
initial density of $dN_{g}/dy$ $\sim$ 1000 or a transport coefficient
$\hat{q} \sim$ 3.5 GeV$^{2}$/fm.

\subsection{Limitations and first test of jet measurement at RHIC in $p+p$ collisions}

There are mainly two arguments in favour of direct jet studies
justifying the fact that the ``leading'' particle physics present some
limitations or biases. Its first main limitation is the fact that for
extreme quenching scenario one observes particle emission
predominantly from the surface. An increase of the in medium path
length, can break the correlation of the leading particle studied with
its original parent parton. For this reason, the $R_{AA}$ exhibits
small sensitivity to the medium properties~\cite{18}. This phenomenon
is denoted as the {\it surface bias} effect. Moreover, using leading
particles for the analysis, one would be biased by the fluctuations
induced by the production spectrum itself. Due to its steeply falling
shape with increasing $E_{T}$, the produced jet with $E_{T}^{\rm prod}$
energy will have fluctuations which will populate with a higher
probability an energy part of the spectrum higher than an average
reconstructed energy $E_{T}^{\rm rec}$. Typically, the mean value of the
distribution of the leading particles for a fixed jet energy is
about 18\%. For a given reconstructed energy interval, the lower
energy jets (mainly produced) that have a harder than average
longitudinal fragmentation will significantly increase the average
energy fraction carried by the leading particles up to $\sim$ 60\%
(section 3.1). It is known as the {\it trigger bias}
effect~\cite{12,18}.

\begin{figure}[htbp]
\setcaptionmargin{5mm}
\onelinecaptionstrue
\includegraphics[scale=0.45, angle=0]{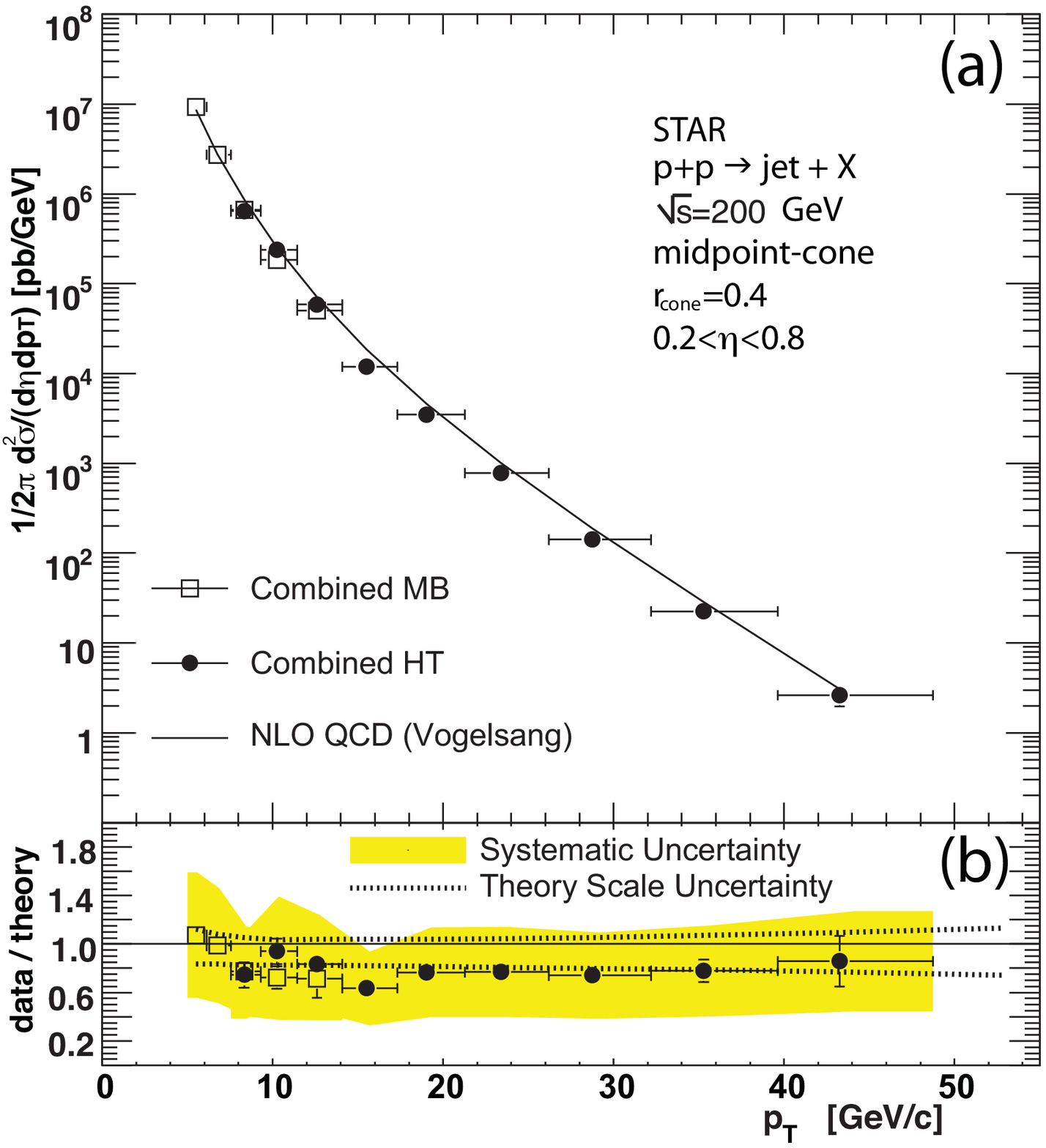}\\
\includegraphics[scale=0.22, angle=0]{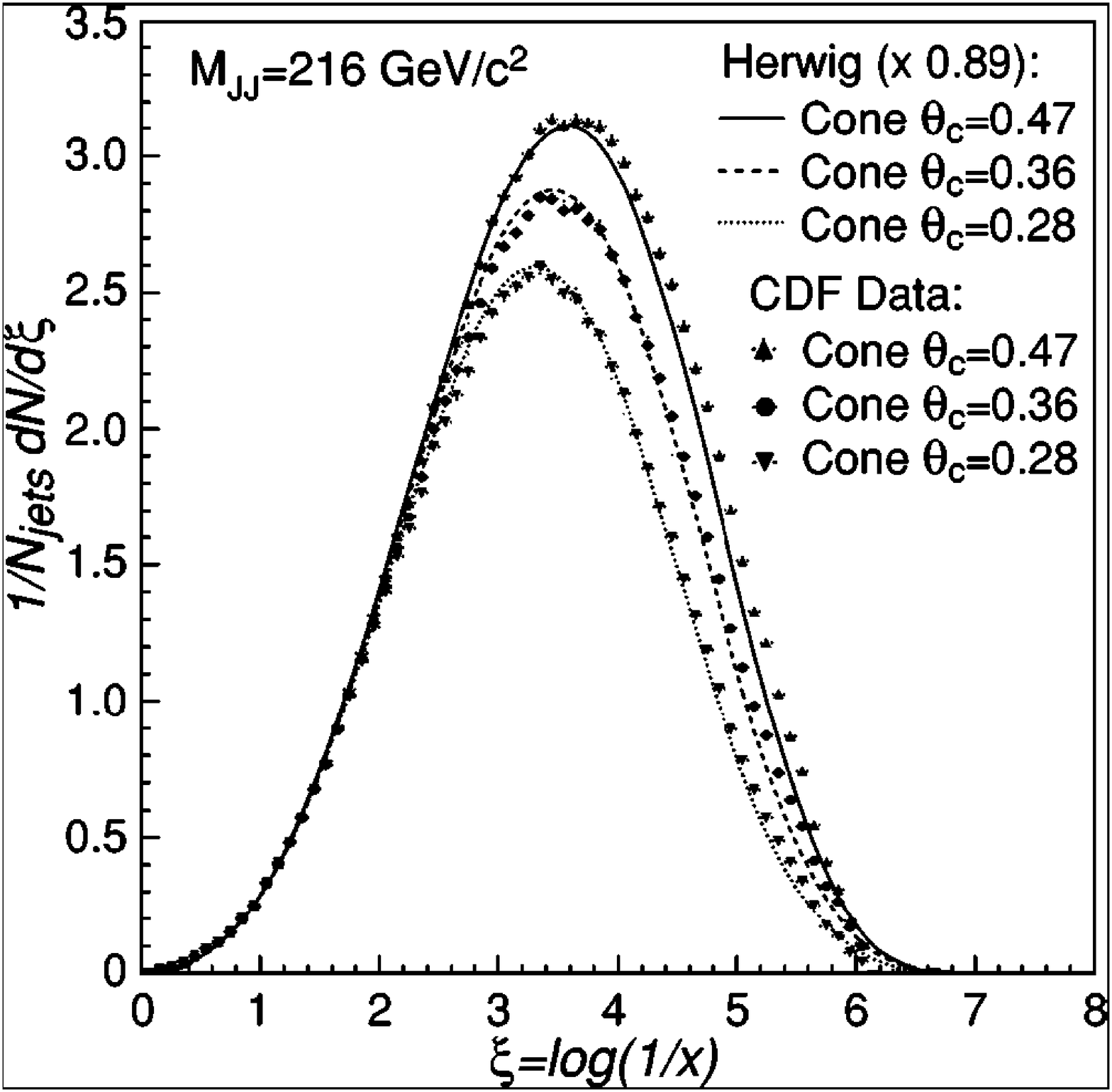}
\includegraphics[scale=0.22, angle=0]{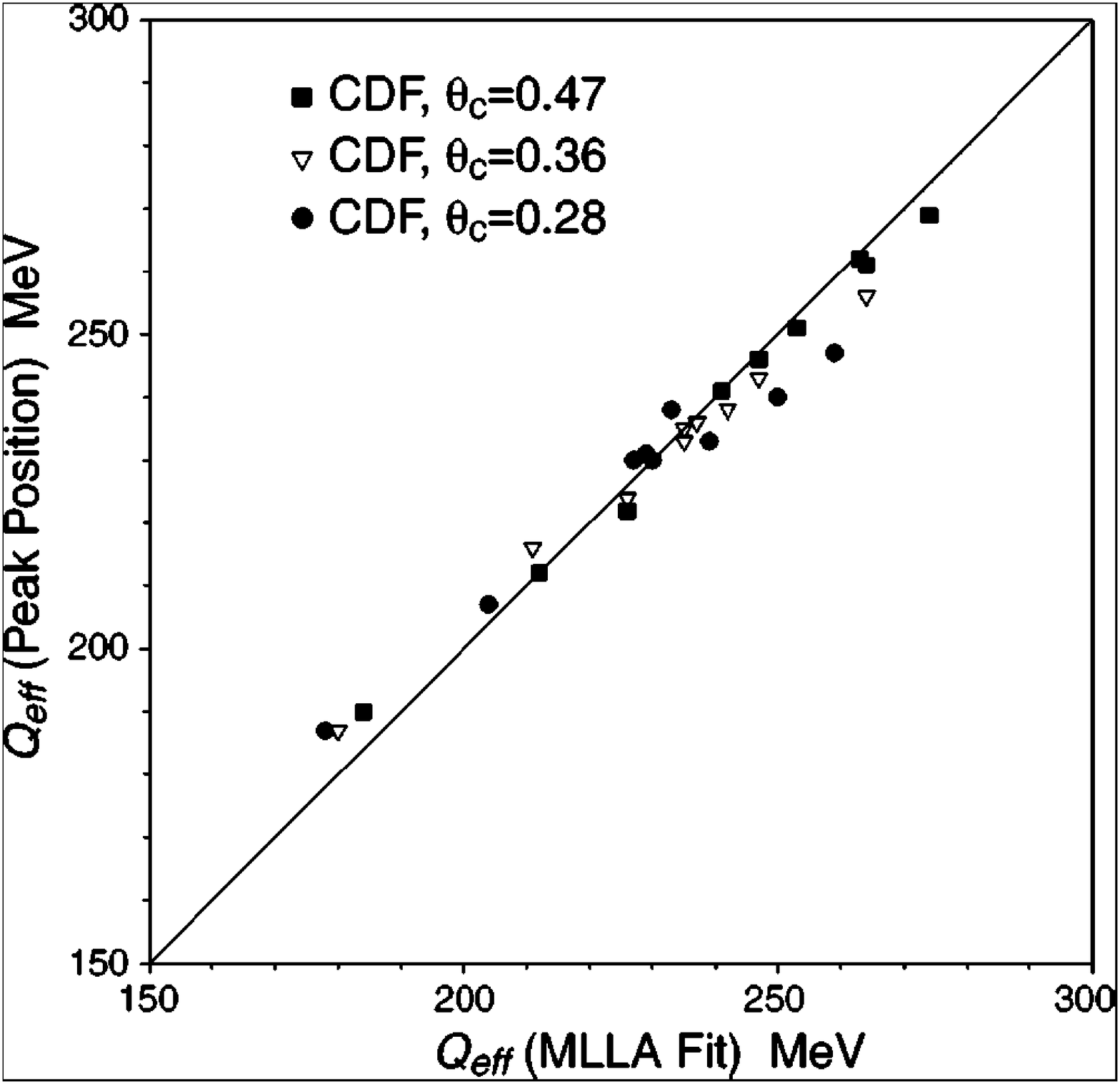}
\caption{{Top:} inclusive differential cross section for $p+p\rightarrow$jet+ X at $\sqrt{s_{NN}}$ = 200 GeV versus jet p$_{T}$ for a jet cone radius of 0.4 (more details in~\cite{19}) {Bottom left:} comparison of the inclusive momentum distribution of particles in jets in the restricted cones of size $\Theta_{C}$ = 0.28, 0.36 and 0.47 rad to HERWIG 5.6 predictions (scaled by 0.89), for dijet mass bin M$_{jj}$ = 216 GeV/$c^{2}$. {Bottom right:} correlation between Q$_{eff}$ extracted from a MLLA fit and from a Gaussian fit for peak position. Uncertainties (not shown) are dominated by the systematic errors~\cite{20}.}
\label{fig:STARjets}
\end{figure}

Ideally, the analysis of reconstructed jets should increase the
sensitivity to medium parameters by reducing the trigger bias. It will
also allow to measure the original parton 4-momentum as well as the
jet structure.
Even if the jet reconstruction is impossible in HIC at RHIC, one can
wonder to which extent jet measurement in experiments such as STAR (in
$p+p/d+$Au) and ALICE (in all systems), which do not use full calorimetry
but an association of charged particles from tracking and neutral
particles from calorimetry, is feasible.
Figure~\ref{fig:STARjets} (top) definitively answers this
question. It presents the first measurement of reconstructed jets
in $p+p$ polarized collisions at $\sqrt{s_{NN}}$ = 200 GeV by the STAR
experiment~\cite{19}. The inclusive differential cross section for the process
$p+p\rightarrow$ jet + X versus jet $p_{T}$ is compared to NLO
calculations. This measurement has been obtained using a midpoint cone
finder algorithm with a cone radius of R = 0.4. Both information from
charged tracks in the Time Projection Chamber (TPC) ( acceptance 0 $< \phi
<$ 2$\pi$ and |$\eta$| $<$ 0.3) and neutrals from lead-scintillator
sampling barrel calorimeter (acceptance 0 $< \phi <$ 2$\pi$ and 0 $<$
|$\eta$| $<$ 1) has been stored in a grid of cell size of $\Delta\eta
\times \Delta\phi$ = 0.05$\times$0.05. We note: 1 - the pure power law shape
of the spectrum in agreement with NLO calculations, 2 - the jet
p$_{T}$ range covered up to 50 GeV/$c$. This measurement is really
promising for the physics foreseen in ALICE.

A natural extension to such measurement would be to study the
capabilities of heavy ion dedicated experiments to make precision
measurements in order to test pQCD ``{\it \`a la } CDF''. The
experimental measured ``Humpbacked plateau'' obtained by the Collider
Detector at Fermilab (CDF) on di-jet events compared to the framework
of the modified leading log approximation (MLLA) and to the hypothesis
of local parton-hadron duality (LPHD) is presented in
Fig.~\ref{fig:STARjets} (bottom-left and bottom-right)~\cite{20}. The distribution presented in the left figure 
is the jet fragmentation function in the variable $\xi$. The evolution
of the distribution has been studied as a function of jet energy and
cone angle $\theta_{c}$ around the jet axis. The MLLA evolution equations
should allow an analytical description of the parton shower for gluon
and quark jets insuring color coherence effect.
The LPHD hypothesis assumes that hadronization is local and occurs at
the end of the partons shower development, so that properties of
hadrons are closely related to those of partons. MLLA+LHPD scheme
views jet fragmentation as a predominantly perturbative QCD process.
A possible study that can be made is illustrated in
Fig.~\ref{fig:STARjets}, right plot. A MLLA fit to these distributions or
the extraction of the peak position from a gaussian fit should allow
to constrain the determination of the phenomenological scale
$Q_{eff}$, the only parameter of the model. It is then interesting to
wonder to which extent such study would be accessible to the ALICE
experiment. It would be necessary to first estimate if, with the
expected luminosity and with the ALICE acceptance, the di-jet rate
would be high enough to perform such measurement.

\section{Jet reconstruction on an event by event basis at LHC energies}
\label{sec:III}

\subsection{ALICE detectors and jet rates}

ALICE is a multipurpose heavy-ion experiment~\cite{21}. Here, we will
concentrate on the detectors we have used to perform jet
reconstruction studies. The central tracking system (here ITS+TPC),
which has good PID capabilities, makes ALICE a dedicated experiment
for heavy ion studies. It covers a full azimuthal acceptance but is
limited to the mid-rapididity region (|$\eta$| < 0.9). Its excellent
momentum resolution for charged particles covers the large range of
100 MeV/$c$ to 100 GeV/$c$~\cite{12}. The capabilities of ALICE to
disentangle particles with good PID down to very low $p_{T}$ should
lead to a very precise measurement of the number of particles inside a
jet, especially at low p$_{T}$ where strong modifications of the
fragmentation function are expected in a HI environment. One can
notice however that the tracking system is quite slow responding. It
would then be necessary to include a detector with excellent trigger
response. This point is one of the two arguments that have motivated
the project to include an electromagnetic calorimeter (EMCal) in the
ALICE apparatus. EMCal should bring trigger capabilities to ALICE and
will also improve the mean reconstructed jet energy as well as its
energy resolution. The proposed detector is an electromagnetic
Pb-scintillator sampling calorimeter with a design energy resolution
of $\Delta E/E$ = 10\%/$\sqrt{E}$ and a radiation length of $\sim$ 20
X$_{0}$. It contains a total of around 13k towers in Shashlik geometry
with a quite high granularity ($\Delta\eta
\times \Delta\phi$ = 0.014 $\times$ 0.014). Its main limitation is
probably its not so extended ($\eta$,$\phi$) acceptance (|$\eta$| < 0.7,
80$^{\circ}$ $ < \phi < $ 190$^{\circ}$)~\cite{7,8}.

\begin{figure}[htbp]
\setcaptionmargin{5mm}
\onelinecaptionstrue
\includegraphics[scale=0.4, angle=0]{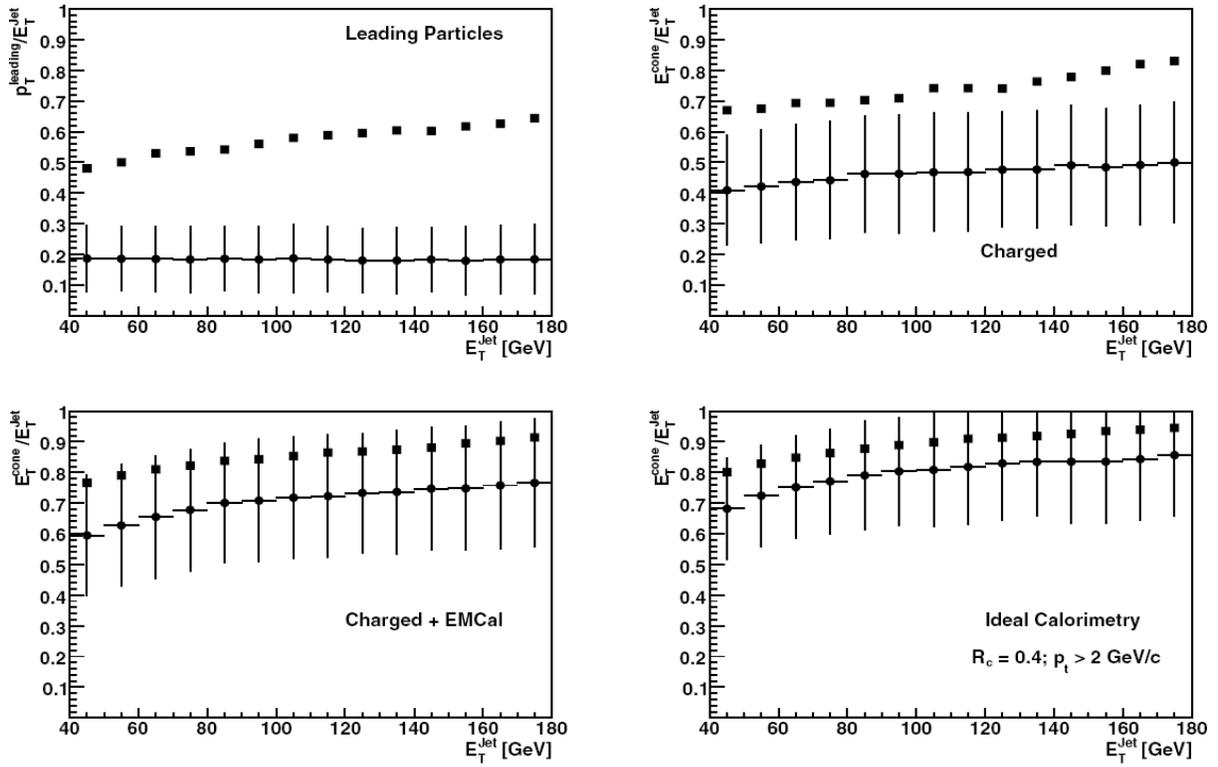}
\caption{The ratio between reconstructed energy and generated energy, E$_{T}^{\hbox{\scriptsize rec}}$/E$_{T}^{\hbox{\scriptsize gen}}$, as a function of the generated energy (circles), for which the RMS values are shown as error bars, and as a function of the reconstructed energy (squares). The former is equivalent to the ratio obtained from monochromatic jets whereas the latter contains the bias induced by the input spectrum~\cite{12}.}
\label{fig:trigger}
\end{figure}

Figure~\ref{fig:trigger} illustrates the improvement expected on the
reconstructed jet energy compared to the generated one using (top
left) leading particles in the reconstruction, (top right) only
charged particles, (bottom left) the association of charged + neutral
particles and (bottom right) an ideal calorimetry case using a cone
radius R = 0.4 and a 2 GeV/$c$ $p_{T}$ cut in Pb+Pb collisions. The
plots present the ratio between the reconstructed energy and the
generated energy as a function of the generated energy [equivalent to
what would be obtained with monochromatic jets as input] (circles) and
as a function of the reconstructed energy [equivalent to what would be
obtained with a full jet spectrum as input] (squares). As discussed
previously, one can see first that for the leading particle case the
ratio is 18\% for monochromatic jets and increases to $\sim$ 60\% due
to the production spectrum bias. Then, the use of reconstructed
charged jets improves the reconstructed energy and reduces the trigger
bias effect. Finally, we reconstruct even higher energy using also the
neutral particles in the reconstruction and the trigger bias is
drastically reduced~\cite{12,18}.

ALICE will study the whole spectrum of jet production from mini-jets
($E_{T} <$ 20 GeV) to high-$E_{T}$ jets of several hundred GeV. From
PYTHIA simulations over a large $\eta$ range of $\pm$ 2.5, one can
note that for $E_{T}$ > 20 GeV, 17\% of the produced jets are in the
ALICE fiducial region |$\eta$| $ < $ 0.5 and in only 8\% of these jet
events the back-to-back jet-pairs are in the acceptance. For jets with
$E_T > $ 100 GeV, the first number increases to 26\% with the same
fraction of di-jets in the ALICE acceptance. With the expected average
luminosity of 5 $\times$ 10$^{\hbox{\scriptsize 26}}$
cm$^{-2}$s$^{-1}$, the number of jets produced per effective months of
running (10$^{6}$s) within the fiducial region |$\eta$|<0.5 for
charged jet reconstructed using ALICE central barrel tracking ITS+TPC
has been estimated. For $E_{T} <$ 100 GeV, the expected jet rate is
high enough even with the limited read-out rate of the TPC so that
more than 10$^{4}$ jets will be measurable. For energy jets higher
than 100 GeV, triggering will be necessary to be able to perform
fragmentation function analysis.

\subsection{Expected $p+p$ performances using calorimetry}
The tools available for jet reconstruction will not be discussed in
this contribution neither the expected physics performances of
ALICE. For really nice discussion on these points one can already
refer to the ALICE PPR II~\cite{12}. Our aim in this last part will be
to concentrate on a study comparison which has been performed on jet
reconstruction in $p+p$ collisions using only charged particles (denoted
as ``C'' in the following) and charged + neutral particles (CN) from
full simulation. For such study, PYTHIA events of monochromatic jets
of 50, 75 and 100 GeV have been produced and passed through the full
detector simulation and reconstruction chain using GEANT3. Jets have
been simulated inside the EMCal acceptance. The analysis framework of
ALICE for jet reconstruction uses a seeded cone finder algorithm based
on the UA1 cone finder algorithm~\cite{22}. The parameters and conditions that
have been used for this study have been chosen in order to reproduce
the jet finder conditions that will be used for the jet reconstruction
in a HI environment. Studies have shown that in Nucleus-Nucleus
collisions, because of the really large amount of background energy
that is expected from the underlying event inside a jet cone as well
as its fluctuations, one has to use experimental tricks to reduce the
background effect in the jet finding. It appears that reducing the
cone size and applying a transverse momentum cut on particles
considerably reduce the background energy and eventually allow to
clearly disentangle the signal energy from the background with a
$\sim$ 40\% resolution for a 100 GeV jet (R = 0.4 and p$_{T}$ cut = 2
GeV/c) using charged particles only and pure MC simulations. A 30\%
resolution is expected from pure MC using also neutrals in the
reconstruction~\cite{23}.

\begin{figure}[htbp]
\setcaptionmargin{5mm}
\onelinecaptionstrue
\includegraphics[scale=0.34, angle=0]{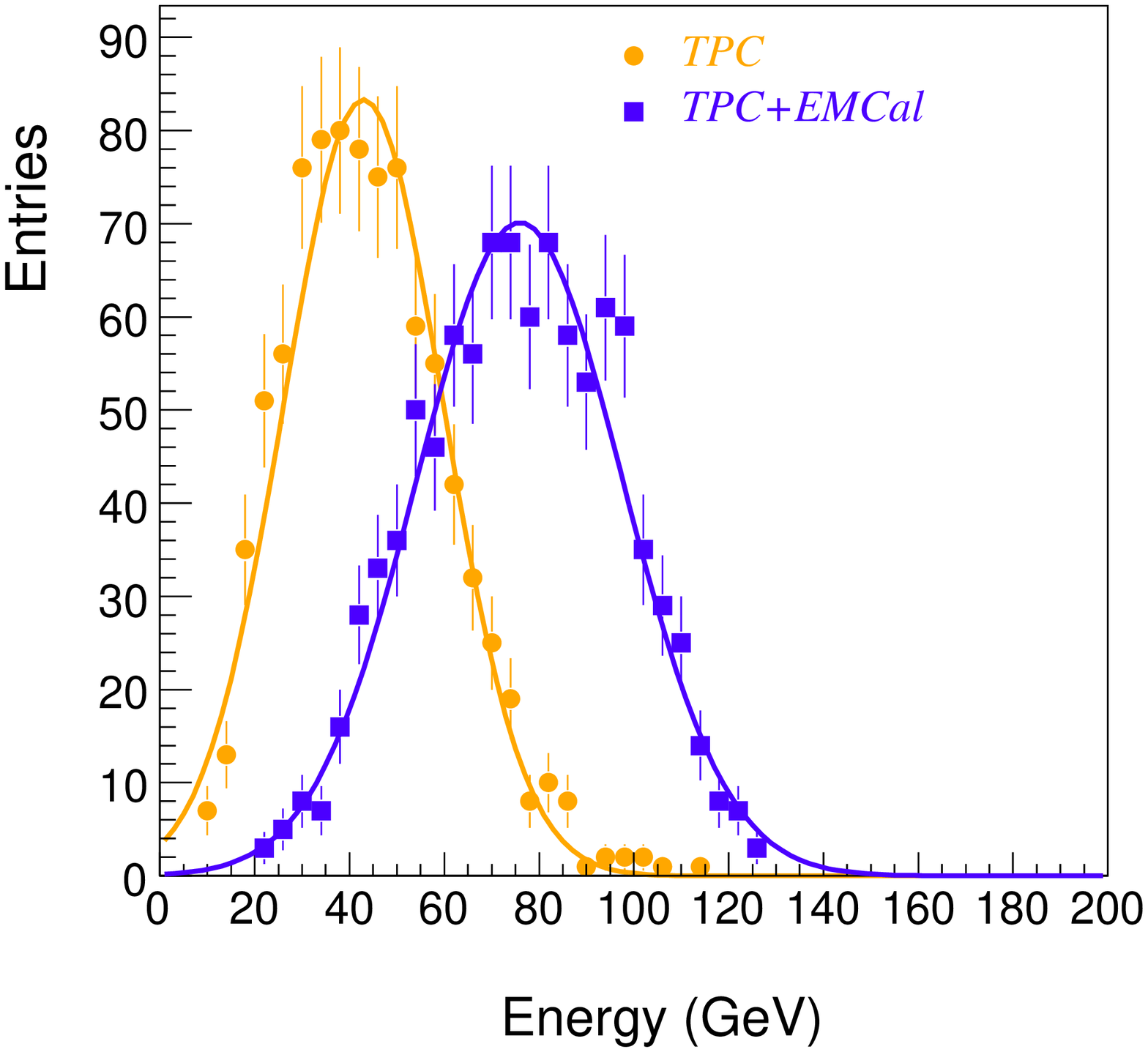}
\includegraphics[scale=0.44, angle=0]{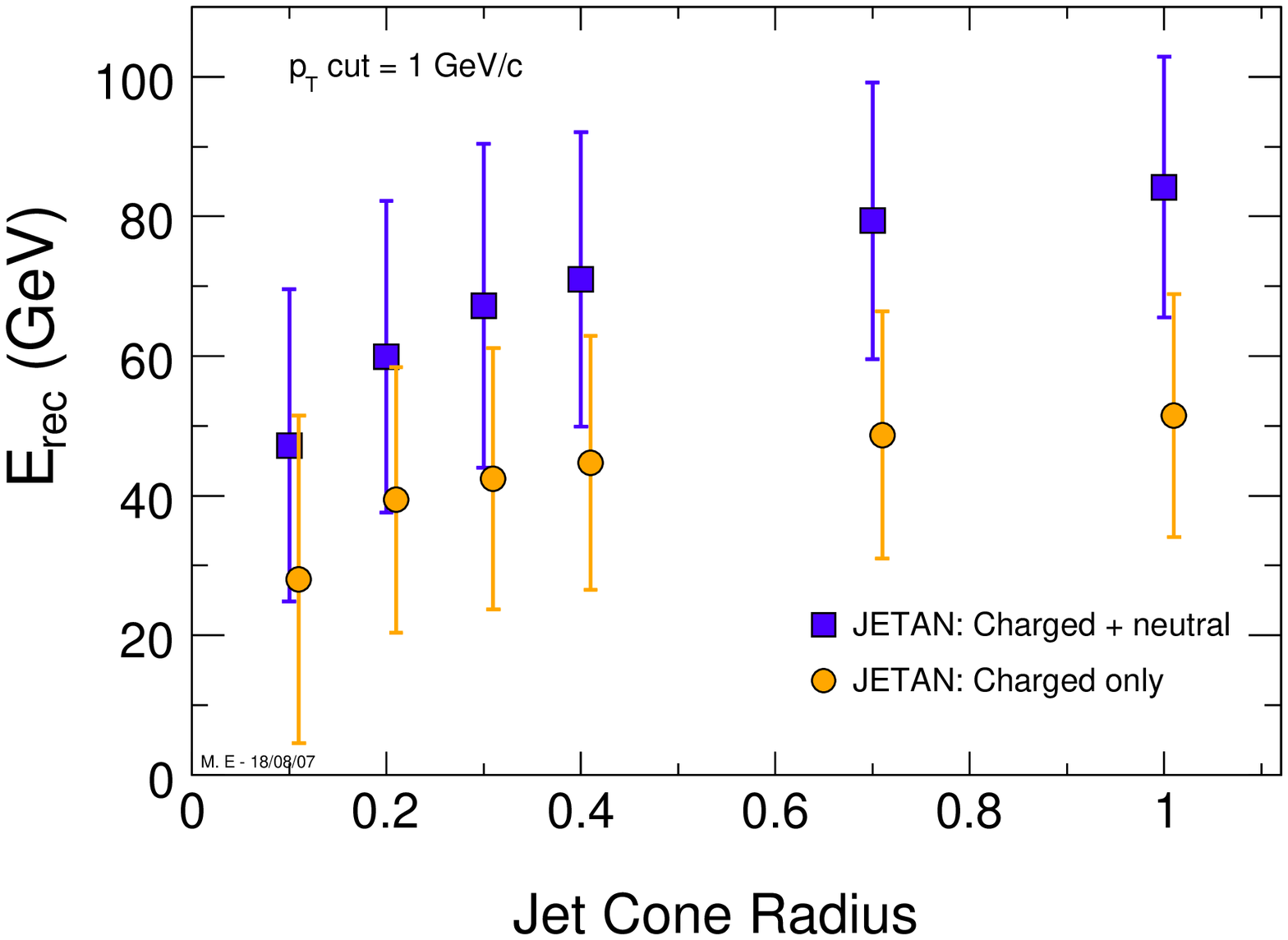}
\caption{{Left:} jet cone energy obtained from the full (PYTHIA + GEANT3) detector simulation of 100 GeV jets in p+p collisions at $\sqrt{s_{NN}}$ = 14 TeV in the EMCal acceptance using charged particles only in the jet finding (circles) and charged + neutral particles, neutrons and $K^{0}_{L}$ excepted (squares). {Right:} jet cone energy as a function of cone sizes for the same cases as fig. left. On this distribution, a p$_{T}$ cut of 1 GeV/$c$ has been applied on tracks. }
\label{fig:Erec}
\end{figure}

Restricting to the $p+p$ case, we show in Fig.~\ref{fig:Erec} (left) the
improvement brought by the calorimeter on the jet cone energy obtained
after jet finding has been applied. Note that the reconstructed jets
which have been kept in such distribution and in the following one are
objects which are completely included in the EMCal acceptance. On this
specific case, R = 0.4 has been used but no p$_{T}$ cut has been
applied on charged particles. The mean is increased from 43 GeV in the
C case to 74 GeV in the CN case. The resolution defined as the ratio
of RMS over Mean is also improved from $\sim$ 40\% to less than
30\%. Figure~\ref{fig:Erec} (right) compares the jet energy within the
cone for varying cone sizes (denoted as $E_{\rm rec}$
distribution) obtained after jet finding in the C and CN cases for 100
GeV jets. Note that a 1 GeV/$c$ $p_{T}$ cut has been applied and that, in
the R = 1 case, part of the reconstructed jet is outside the EMCal
acceptance. The error bars are the RMS of the energy
distributions. $E_{\rm rec}$ increases with increasing
R and seems to saturate around 84 GeV. Such value can be explained by
detector reconstruction efficiency, the $p_{T}$ cut applied, the
neutral particles missed in the reconstruction outside the EMCal
acceptance as well as the non measured neutrons and K$_{L}^{0}$.

\begin{figure}[htbp]
\setcaptionmargin{5mm}
\onelinecaptionstrue
\includegraphics[scale=0.34, angle=0]{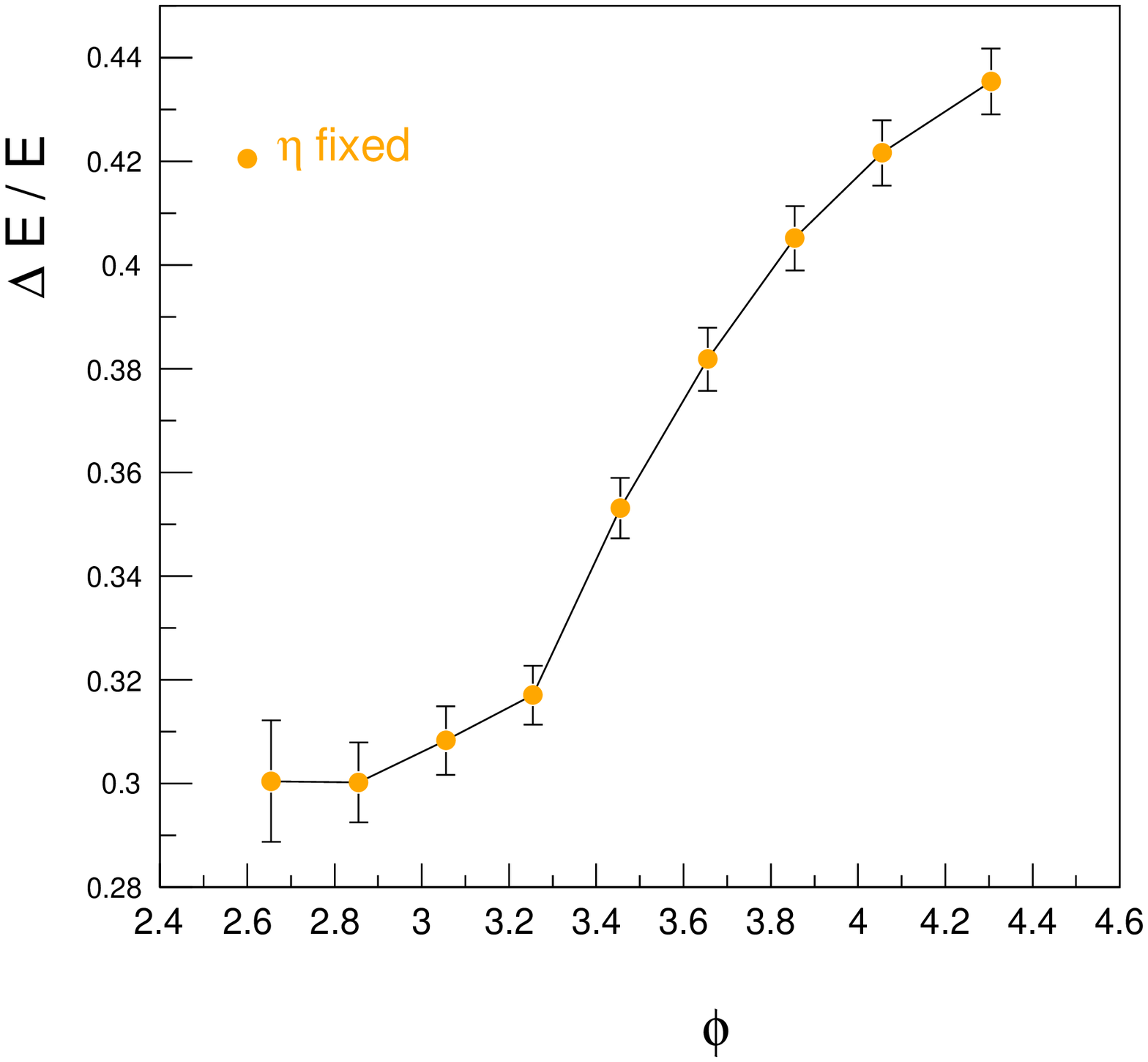}
\includegraphics[scale=0.45, angle=0]{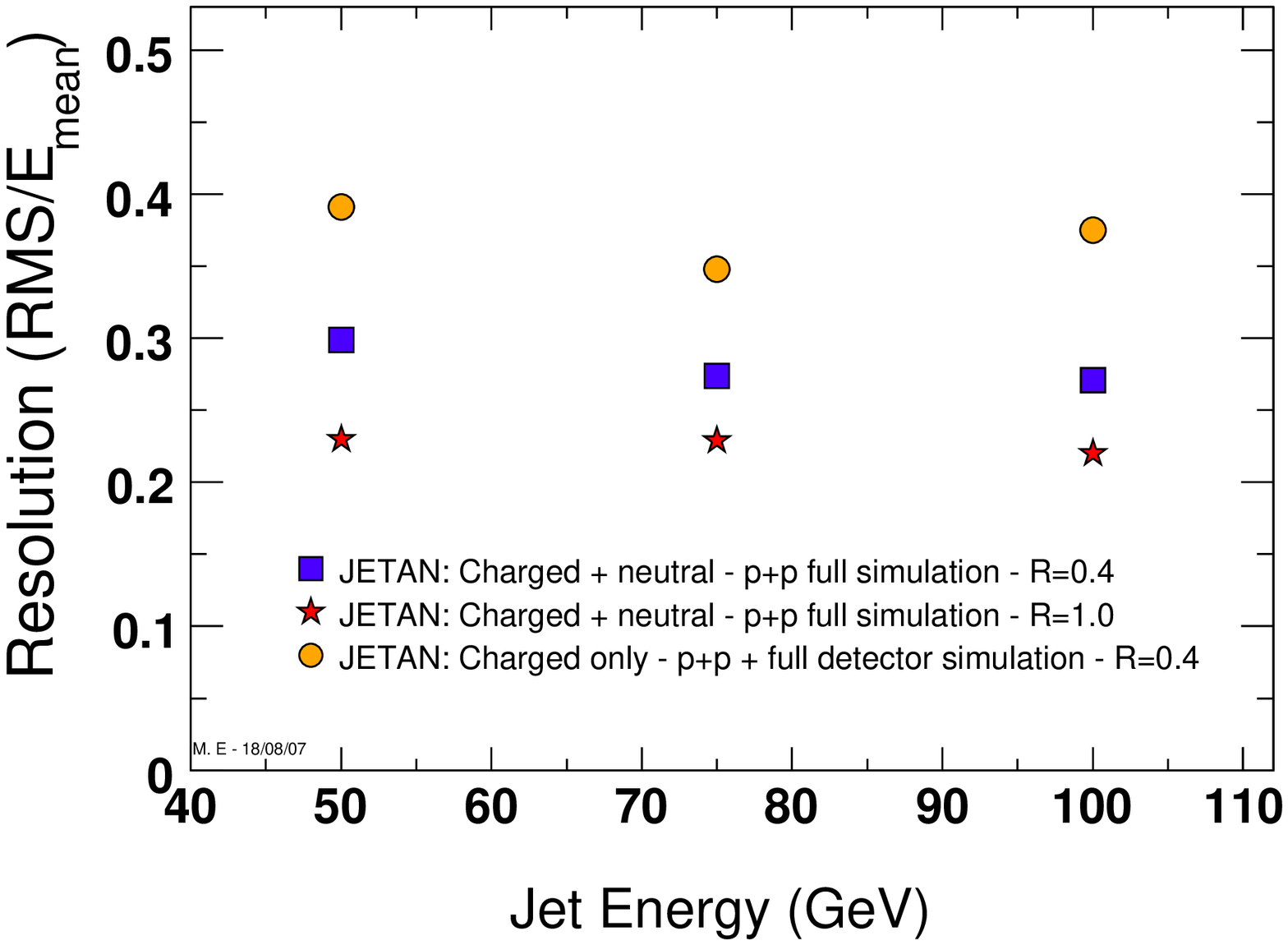}
\caption{{Left:} jet energy resolution (RMS/E$_{\rm mean}$) as a function of the maximum limit position of the jet center in the $\phi$ direction. {Right:} jet energy resolution of monoenergetic jets of 50, 75 and 100 GeV using only charged particles with a cone radius R = 0.4 (circles), using charged and neutral particles with R = 0.4 (squares) and R = 1.0 (stars). In the case of R = 1.0, part of the jet is outside the EMCal acceptance.}
\label{fig:reso}
\end{figure}

The previous figures included only jets which were selected on
geometrical criteria in order not to be biased by the EMCal edge effects
on the reconstructed jet energy. Only jets whose center was chosen in
a given ($\eta,\phi$) window assuring all the jets to be included in
the calorimeter were kept. Keeping $\eta$ fixed, one can then open the
$\phi$ accessible window to the center of jet and study how the
resolution evolves consequently. This is presented in
Fig.~\ref{fig:reso} (left). $\phi\sim$ 3.3 rad corresponds to the
calorimeter edge position in $\phi$. Fig.~\ref{fig:reso} suggests that as long as
the center of the jet is taken inside the EMCal acceptance, the
resolution remains acceptable. It quickly becomes worse with the jet
center taken outside the calorimeter. In this distribution, the first
point is the limit to which a jet is completely included in the
calorimeter. The last point tends to C case even if the energy
distribution is a bit distorted by the neutral contribution.  We
eventually end-up this discussion by showing the resolutions obtained
for different monoenergetic jet energies in three different cases: C
with R = 0.4 (circles), CN with R = 0.4 (rectangles) and CN with R = 1
(stars). Even if the resolution is a bit worse for 50 GeV jets, we
obtain a resolution lower than 40\% in the C case which is really
improved by the inclusion of the neutrals in the jet finding
procedure. The resolution with R = 1 is a bit worse than expected for
the geometrical reasons already discussed above.

\section{Conclusion and perspectives}
Starting from a discussion on RHIC results, we have tried to motivate
the necessity to move from leading particle to jet physics to extract
more information on the jet quenching phenomenon. The ALICE experiment,
with its tracking and PID capabilities, will be very well equipped for
such studies. A good improvement in the jet reconstruction is expected
from the inclusion of the EMCal calorimeter in the ALICE
apparatus. Some results obtained from full simulation in $p+p$ have been
provided to show the modifications engendered by the inclusion of
neutral particles in jet finding. In a near future, we
will show the improvement brought by the calorimeter on Pb+Pb
collisions as well as the expected fragmentation function and its
modification in a HI environment with full simulation.

\newpage
%

\newpage

\newpage

\newpage

\newpage

\end{document}